\documentclass[12pt,reqno]{amsart}
\usepackage{amsmath,amssymb,amsfonts,amsthm}
\usepackage[mathscr]{eucal}
\usepackage[all]{xy}
\usepackage{hyperref}
\usepackage{setspace}
\usepackage{color}
\allowdisplaybreaks

\author{V.~A.~Abakumova and S.~L.~Lyakhovich}

\address{Physics Faculty, Tomsk State University, Lenin ave. 36, Tomsk 634050, Russia.}

\email{abakumova@phys.tsu.ru, sll@phys.tsu.ru}
\allowdisplaybreaks

\title{Reducible Stueckelberg symmetry and dualities}

\begin{document}

\maketitle
\begin{abstract}
We propose a general procedure for iterative inclusion of
Stueckelberg fields to convert the theory into gauge system being
equivalent to the original one. In so doing, we admit reducibility
of the Stueckelberg gauge symmetry. In this case, no pairing exists
between Stueckelberg fields and gauge parameters, unlike the
irreducible Stueckelberg symmetry. The general procedure is
exemplified by the case of Proca model, with the third order
involutive closure chosen as the starting point. In this case, the
set of Stueckelberg fields includes, besides the scalar, also the
second rank antisymmetric tensor. The reducible Stueckelberg gauge
symmetry is shown to admit different gauge fixing conditions. One of
the gauges reproduces the original Proca theory, while another one
excludes the original vector and the Stueckelberg scalar. In this
gauge, the irreducible massive spin one is represented by
antisymmetric second rank tensor obeying the third order field
equations. Similar dual formulations are expected to exist for the
fields of various spins.
\end{abstract}

\section{Introduction}
Since the original Stueckelberg's work \cite{Stueckelberg}, the idea
remains attractive for decades  concerning inclusion of auxiliary
fields into the action in such a way that modified theory becomes
gauge invariant while it is still equivalent to the original one.
The reviews and further references can be found in
\cite{Ruegg:2003ps}, \cite{Boulanger:2018dau}.

In the constrained Hamiltonian formalism, the Stueckelberg idea has
transformed into a method of converting the second class constraints
into the first class ones \cite{Faddeev:1986pc},
\cite{Batalin:1986aq}. The conversion is achieved by extending the
phase space by extra dimensions, that can be understood as
introduction of Stueckelberg fields. The local existence theorem for
the conversion procedure has been proven in the article
\cite{Batalin:1991jm}, the global proof of the conversion existence
can be found in \cite{Batalin:2005df}. The starting point of the
Hamiltonian conversion is a complete system of the constraints
including primary and secondary ones of all the generations. The
conversion variable is assigned to every second class constraint.
Given the complete system of constraints, the Hamiltonian conversion
works as a systematic iterative procedure which is proven
unobstructed. Unlike the Hamiltonian counterpart, the common
practice of including Stueckelberg fields in Lagrangian formalism
seems more art than science. Most often this works as a
``Stueckelberg trick'', which implies that the action is split into
gauge and non-gauge parts. The Stueckelberg gauge symmetry of the
original fields is assumed to remain the same as for the gauge
invariant part of the action, while the transformations of
Stueckelberg fields are chosen to compensate the non-invariance of the
rest part. The choice of this split into gauge and non-gauge
parts is an art, and it can be ambiguous. It is even unclear, why
such a split is always possible. From the Hamiltonian perspective,
this would mean to assume each second class constraint to be
decomposed into the first class part and the ``symmetry breaking
part". The Hamiltonian conversion method proceeds from any complete
set of constraints, not assuming the possibility of any decomposition of
the constraints.

Recently, a systematic procedure has been proposed for covariant
inclusion of the Stueckelberg fields \cite{Lyakhovich:2021lzy} in
Lagrangian formalism.  The starting point for the method is the
\emph{involutive closure} of the original Lagrangian system of field
equations. The original equations can be non-involutive, i.e. they
can admit the lower order consequences. Completion of the system of
the field equations by their consequences is understood as an
involutive closure, if the completed system does not admit any
further lower order consequences. In principle, the involutive
closure can include also the higher order consequences. Completion
of the Hamiltonian constrained system by the secondary constraints
is an example of the involutive closure. The involutively closed
form of the field equations allows one to count the degree of
freedom number in an explicitly covariant manner \cite{Involution}.
The procedure of the article \cite{Lyakhovich:2021lzy} allows one to
iteratively include Stueckelberg fields for any field theory
proceeding from the involutive closure of the original Lagrangian
equations, and it is proven to be unobstructed. This procedure
implies inclusion of independent consequences into the involutive
closure of Lagrangian equations. Given this starting point, one
arrives at the irreducible Stueckelberg gauge symmetry.

In this article,  we consider inclusion of Stueckelberg fields
proceeding from the involutive closure which involves a reducible
set of consequences of Lagrangian equations. This leads to two main
distinctions from the case of independent consequences. First, there
is no pairing anymore between the Stueckelberg fields and gauge
parameters. Second, the Stueckelberg gauge symmetry turns out
reducible. There are no obstructions to inclusion of the
Stueckelberg fields in the reducible case, much like to the
irreducible one. To exemplify the general procedure, we consider the
third order involutive closure of the Proca equations when the
original equations are also complemented, besides the first order
consequence, by the antisymmetric combinations of the derivatives of
the Lagrangian equations. This leads to inclusion, besides the usual
Stueckelberg scalar, of the Stueckelberg field, being the second
rank antisymmetric tensor.  Full Stueckelberg symmetry mixes the
original vector with all the Stueckelberg fields. This reducible
gauge symmetry admits different gauge fixing conditions. The
simplest gauge kills all the Stueckelberg fields reducing the
dynamics to the original Proca equations. The alternative gauge
fixing condition is also admissible such that kills the Stueckelberg
scalar \emph{and the original vector field}, while all the dynamics
is described by the antisymmetric tensor $B^{\mu\nu}$ obeying the
third order equation,
\begin{equation}\label{B}
    (\Box +m^2)\partial_\nu B^{\mu\nu}=0 \, ,
\end{equation}
with appropriate gauge fixing for gauge symmetry\footnote{For the
details of gauge symmetry and gauge fixing of the Stueckelberg
field, see in the Section 3.} of $B^{\mu\nu}$. By itself, this
non-Lagrangian equation, being the gauge fixed form of the reducible
Stueckelberg Lagrangian system, describes the irreducible massive
spin 1, much like the original Proca equation. The reason is
obvious: the Proca model is equivalent to the Klein-Gordon equation
supplemented by the transversality condition,
\begin{equation}\label{KG}
(\Box +m^2)A^\mu=0\, ,\qquad \partial_\mu A^\mu =0 \, .
\end{equation}
In Minkowski space, any transverse vector is a divergence of the
antisymmetric tensor,
\begin{equation}\label{Transverse}
\partial_\mu A^\mu=0\quad\Leftrightarrow\quad
\exists\,B^{\mu\nu}=-B^{\nu\mu}: \, A^\mu=\partial_\nu B^{\mu\nu}.
\end{equation}
In a sense, $B^{\mu\nu}$ is a ``potential" for the transverse vector
$A^\mu$. The non-Lagrangian equations (\ref{B}) can be viewed as a
reformulation of the Proca model in terms of the potential, such
that automatically accounts for the transversality condition. Under
the proposed procedure of inclusion of the Stueckelberg fields, both
dual formulations, (\ref{B}) and (\ref{KG}), are included into a
uniform Lagrangian theory even though one of them is non-Lagrangian
by itself. Imposing appropriate gauge fixing conditions, one can
switch from the vector formulation to the dual one, and vice versa.
As explained in the conclusion, it seems to be a general phenomenon
which extends to other representations and goes beyond the free
level.

The article is organized as follows. In the next section, the
general scheme of inclusion of the Stueckelberg fields is outlined for
the case of reducible Stueckelberg gauge symmetry. In Section 3, the
general procedure is exemplified by unconventional inclusion of
Stueckelberg fields in the Proca model such that leads to reducible
gauge symmetry. The results and further perspectives are discussed
in the Conclusion.

\section{Inclusion of Stueckelberg fields with reducible gauge symmetry}

As a preliminary, let us explain the strategy of including
Stueckelberg fields implemented in this Section. First, the
Lagrangian equations are complemented by the consequences such that
the entire system is involutive. Once the completed system is
non-Lagrangian, the second Noether theorem does not apply, and the
gauge identities arise, being unrelated to the gauge symmetry. The
general structure  of gauge algebra is known for not necessarily
Lagrangian field equations \cite{Lyakhovich:2004xd},
\cite{Kazinski:2005eb}. For the case when the non-Lagrangian system
is a completion of the Lagrangian one, the gauge algebra has some
specifics which are detailed as the second step. As the third step,
we introduce the Stueckelberg fields with two goals. First, the
involutive system should be zero order of the expansion of
Lagrangian Stueckelberg equations. Second, the gauge identities of
the involutive closure of the original system should be reproduced
as zero order (in Stueckelberg fields) of Noether identities for
Stueckelberg action. This defines zero order of gauge symmetry
generators and the first order of the action. The existence of all
the higher orders can be proven along the similar lines to the
irreducible case \cite{Lyakhovich:2021lzy}.

In this Section, we use the condensed notation. All the condensed
indices are supposed to include numerical labels and the
space-time points. Summation over the condensed index implies
integration over $x$. The partial derivatives are understood as
variational.

Consider a theory of fields $\phi^i$ with the action $S(\phi)$.
Lagrangian equations read
\begin{equation}\label{LE}
\partial_i S(\phi)=0 \, .
\end{equation}
In this article, we consider a theory where the original action
does not have gauge symmetry. This means that any identity between
the field equations (\ref{LE}) has a trivial generator which
vanishes on shell
\begin{equation}\label{trivR}
\kappa^i\partial_i S\equiv 0 \quad \Leftrightarrow \quad \exists\,
E^{ij}=-E^{ji}: \quad \kappa^i=E^{ij}\partial_jS \, .
\end{equation}
Inclusion of the Stueckelberg fields in the gauge invariant actions
will be considered elsewhere.

Let us complement the field equations (\ref{LE}) by their
differential consequences,
\begin{equation}\label{tau}
\tau_\alpha(\phi)=-\Gamma_\alpha{}^i(\phi)\partial_i S(\phi) \, ,
\end{equation}
where $\Gamma_\alpha{}^i(\phi)$ are supposed to be  local
differential operators. The generators $\Gamma$ of the consequences
are considered equivalent if they lead to the same $\tau$. Hence,
the equivalence relation reads
\begin{equation}\label{equivG}
\Gamma_\alpha{}^i\,\sim\,
\Gamma_\alpha^\prime{}^i\quad\Leftrightarrow\quad \Gamma_\alpha{}^i-
\Gamma_\alpha^\prime{}^i = E_\alpha{}^{ij}\partial_j S, \quad
E_\alpha{}^{ij}=-E_\alpha{}^{ji}\, .
\end{equation}
The completed system
\begin{equation}\label{Closure}
\partial_i S(\phi)=0\,, \quad \tau_\alpha(\phi)=0
\end{equation}
is assumed involutively closed, i.e. all the lower order
consequences are already contained among equations (\ref{Closure}).
Obviously, the involutive closure (\ref{Closure}) is equivalent to
the original system, because all their solutions coincide. By
construction, the involutively closed system enjoys gauge identities
\begin{equation}\label{GI}
\Gamma_\alpha{}^i(\phi)\partial_i S(\phi)+\tau_\alpha(\phi)\equiv
0\, ,
\end{equation}
while there are no gauge symmetry. Let us assume the set of the
generators $\Gamma_\alpha{}^i$ of consequences (\ref{tau})  is
over-complete,
\begin{equation}\label{Z}
Z_A{}^\alpha\Gamma_\alpha{}^i=E_A{}^{ij}\partial_jS\,, \quad
E_A{}^{ij}=-E_A{}^{ji}\,,
\end{equation}
i.e. certain combinations of $\Gamma$'s reduce to the trivial gauge
generators (\ref{trivR}). This results in the identities between the
consequences (\ref{tau}):
\begin{equation}\label{Ztau}
Z_A{}^\alpha\tau_\alpha\equiv 0 \, .
\end{equation}
The generators of identities are considered equivalent if they
differ by the trivial generator vanishing on shell,
\begin{equation}\label{Zequiv}
Z_A{}^\alpha\,\sim\, Z_A^\prime{}^\alpha \quad\Leftrightarrow\quad
Z_A{}^\alpha\,-\, Z_A^\prime{}^\alpha =E_A{}^{\alpha\beta}\tau_\beta\,
,\quad E_A{}^{\alpha\beta}=-E_A{}^{\beta\alpha} \, .
\end{equation}
 The operators $Z_A{}^\alpha$  are assumed to constitute the generating set for the
 null-vectors of the consequences $\tau_\alpha$, i.e. $Z^\alpha\tau_\alpha\equiv 0\,\Leftrightarrow\,Z^\alpha=\zeta^A Z_A{}^\alpha$.
The identities (\ref{Ztau}) can admit further reducibility,
\begin{equation}\label{Z1}
\exists\, Z_{1A_1}{}^A: \,\, Z_{1A_1}{}^A
Z_A{}^\alpha=E_{A_1}{}^{\alpha\beta}\tau_\beta\,, \,\,
E_{A_1}{}^{\alpha\beta}=-E_{A_1}{}^{\beta\alpha}\, ,
\end{equation}
i.e. certain combinations of the identity generators $Z_A{}^\alpha$
reduce to the trivial null-vectors (\ref{Zequiv}). In principle, the
generating set of the second null-vectors $Z_{1A_1}{}^A$ can be
over-complete in its own turn. In this article, we do not consider
this option assuming no further reducibility.

The set of identities (\ref{GI}), (\ref{Ztau})
between the equations of involutive closure (\ref{Closure}) is assumed complete. This
means, any set of identities, labeled by some condensed index $I$,
reduces to the linear combination of identities (\ref{GI}),
(\ref{Ztau}),
\begin{equation}\label{Lambda}
\begin{array}{c}
\Lambda_I{}^i\partial_iS+\Lambda_I{}^\alpha\tau_\alpha\equiv 0 \quad \Leftrightarrow \quad \exists\, U_I{}^\alpha,\,U_I{}^A:\\
\Lambda_I{}^i\partial_iS+\Lambda_I{}^\alpha\tau_\alpha\equiv
U_I{}^\alpha\big(\Gamma_\alpha{}^i\partial_i
S+\tau_\alpha\big)+U_I{}^AZ_A{}^\alpha\tau_\alpha\,.
\end{array}
\end{equation}
Hence, the generators $\Lambda_I{}^i, \Lambda_I{}^\alpha$ of any
identity between the equations of the system (\ref{Closure}) reduce
to the linear combinations of the generators $\Gamma$ and $Z$ modulo
trivial generators:
\begin{equation}\label{U}
\begin{array}{l}
\Lambda_I{}^i=U_I{}^\alpha\Gamma_{\alpha}{}^i+E_I{}^{ij}\partial_jS+E_I{}^{i\alpha}\tau_\alpha\,, \quad E_I{}^{ij}=-E_I{}^{ji}\,, \\[2mm]
\Lambda_I{}^\alpha=U_I{}^\alpha+U_I{}^AZ_A{}^\alpha-E_I{}^{i\alpha}\partial_iS+E_I{}^{\alpha\beta}\tau_\beta\,,
\quad E_I{}^{\alpha\beta}=-E_I{}^{\beta\alpha}\,.
\end{array}
\end{equation}
Relation (\ref{Z}) leads to the identities between the identities
(\ref{GI}), (\ref{Ztau}), because certain combination of the
identity generators is trivial.

Also notice that the set of the identity generators $Z_A{}^\alpha$  is
over-complete (\ref{Z1}). This leads  to further identities
between the identities (\ref{Ztau}). These second level identities
are irreducible, as their generators $Z_{1A_1}{}^A$ are assumed
independent. Any set of identities, being labeled by the condensed
index $I_1$, between the identities of identities is supposed
generated by $Z_{1A_1}{}^A$:
\begin{equation}
\Lambda_{I_1}{}^AZ_A{}^\alpha\equiv
E_{I_1}{}^{\alpha\beta}\tau_\beta\,, \,\,
E_{I_1}{}^{\alpha\beta}=-E_{I_1}{}^{\beta\alpha}
\quad\Leftrightarrow\quad
\Lambda_{I_1}{}^{A}=U_{I_1}{}^{A_1}Z_{1A_1}{}^A\,.
\end{equation}

Even though original action has no gauge symmetry, the involutive
closure (\ref{Closure}) of Lagrangian equations, being a
non-Lagrangian system, enjoys non-trivial gauge algebra as
demonstrated above. The general idea of inclusion of the Stueckelberg
fields is to cast this gauge algebra back into Lagrangian setup by
introducing extra fields. Specifically, the equations of the
involutively closed system (\ref{Closure}) should be zero order in
the Stueckelberg fields of the Lagrangian Stueckelberg equations,
while the gauge identities (\ref{GI}), (\ref{Ztau}) should be zero
order of Noether identities for the Stueckelberg action. These
reasons lead one to introduce the Stueckelberg field $\xi^\alpha$
for every consequence $\tau_\alpha$ included into involutive closure
(\ref{Closure}), while every gauge identity (\ref{GI}), (\ref{Ztau})
is assigned with the gauge parameter $\epsilon^\alpha, \epsilon^A$.
Given the set of Stueckelberg fields and gauge parameters, we seek
for the Stueckelberg action, and its gauge symmetry, as the power
series in $\xi$:
\begin{equation}\label{S-St}
  \mathcal{S}_{St}(\phi,\xi)=   \sum_{k=0} \mathcal{S}_k\,,\quad
  \mathcal{S}_0(\phi)=S(\phi)\,, \quad
  \mathcal{S}_k(\phi,\xi)=  W_{\alpha_1\ldots\alpha_k}(\phi)\, \xi^{\alpha_1} \ldots
  \xi^{\alpha_k}\,, \quad k>0\,,
\end{equation}
where the first order is defined by the completion functions
(\ref{tau})
\begin{equation}\label{W}
    W_\alpha(\phi)=\frac{\partial \mathcal{S}_{St}(\phi,\xi)}{\partial
    \xi^\alpha}\Big|_{\xi=0}=\tau_\alpha\,.
\end{equation}
Once the gauge identities (\ref{GI}), (\ref{Ztau}) are to be
converted into the Noether identities of the action (\ref{S-St}),
corresponding  gauge parameters $\epsilon^\alpha$ and $\epsilon^A$
are introduced
\begin{equation}\label{GT}
\delta_\epsilon \phi^i=R^i{}_\alpha(\phi,\xi)\epsilon^\alpha+
R^i{}_A (\phi,\xi)\epsilon^A\,, \quad \delta_\epsilon
\xi^\alpha=R^\alpha{}_\beta(\phi,\xi)\epsilon^\beta+R^\alpha{}_A(\phi,\xi)\epsilon^A\,
.
\end{equation}
The gauge symmetry of the Stueckelberg action is equivalent to the
Noether identities between the equations,
\begin{equation}\label{NI}
\delta_\epsilon\mathcal{S}_{St}\equiv 0\,, \,\, \forall\,
\epsilon^\alpha,\, \epsilon^A \, .
\end{equation}
Let us expand the action (\ref{S-St}) and gauge generators
(\ref{GT}) in the Stueckelberg fields $\xi$, and substitute the
expansions into the Noether identities. Comparing the identities
(\ref{NI}) in zero order w.r.t. $\xi$ with the identities (\ref{GI}),
(\ref{Ztau}), we find zero order of the Stueckelberg gauge
transformations,
\begin{equation}\label{GT0}
\delta_\epsilon
\phi^i=\Gamma^i{}_\alpha(\phi)\epsilon^\alpha+\ldots\, ,  \qquad
\delta_\epsilon
\xi^\alpha=\epsilon^\alpha+Z^\alpha{}_A(\phi)\epsilon^A+\ldots \, ,
\end{equation}
where $\Gamma^i{}_\alpha$ are the generators of consequences of
Lagrangian equations (\ref{tau}) included into the involutive
closure of original system, and $Z^\alpha{}_A$ are the generators of
the identities (\ref{Ztau}) between $\tau_\alpha$. The dots stand
for the $\xi$-depending terms. The generators $\Gamma^i{}_\alpha$ of
the consequences (\ref{tau}) are reducible in the sense of relations
(\ref{Z}). This results in the reducibility of the gauge identities
(\ref{GI}), (\ref{Ztau}). Hence, the Noether identities (\ref{NI})
of the Stueckelberg action should be reducible as they begin with
the identities between the equations of the involutive closure.
Reducibility of the Noether identities means the gauge symmetry of
the gauge symmetry. Comparing zero order of identities
(\ref{NI}) with corresponding identities in the system
(\ref{Closure}), we find the gauge transformations of gauge
parameters in zero order w.r.t. Stueckelberg fields
\begin{equation}\label{omega}
\delta_\omega\epsilon^\alpha=Z^\alpha{}_A(\phi)\omega^A+\ldots\, ,
\qquad
\delta_\omega\epsilon^A=-\omega^A+Z_1{}^A{}_{A_1}(\phi)\omega^{A_1}+\ldots
\, ,
\end{equation}
where dots stand for the $\xi$-depending terms, $Z^\alpha{}_A$
are the null-vectors for the generators of consequences (\ref{Z}),
and $Z_1{}^A{}_{A_1}$ are the generators of reducibility for
$Z^\alpha{}_A$, see (\ref{Z1}). The gauge parameters of symmetry for
symmetry are denoted $\omega^A$  and $\omega^{A_1}$. The gauge identities
of identities in the original system (\ref{Closure}) are reducible
again. At the level of Stueckelberg theory this leads to the gauge
symmetry of the parameters $\omega$ from the transformation above.
This symmetry of symmetry in zero order in $\xi$ is generated by the
same operators as in the corresponding identities of identities of
the original system. Hence, the next level gauge symmetry reads
\begin{equation}\label{eta}
\delta_\eta\omega^A=Z_1{}^A{}_{A_1}(\phi)\eta^{A_1}+\ldots \, ,
\qquad \delta_\eta\omega^{A_1}=\eta^{A_1}+\ldots\, .
\end{equation}
The second order in $\xi$ of the Stueckelberg action (\ref{S-St}),
and the first order of the gauge transformations (\ref{GT}), can be
found from Noether identities (\ref{NI}) at the first order, given
the previous order (\ref{W}), (\ref{GT0}). Once the previous order
is found, it is substituted into the expansion of the Noether
identity up to the next order. This allows one to find the next
order, etc. In this way, all the orders of the action and gauge
generators are iteratively found. Up to the second order in $\xi$,
the Stueckelberg action reads
\begin{equation}
\mathcal{S}_{St}=S(\phi)+\tau_\alpha(\phi)\xi^\alpha+\frac{1}{2}W_{\alpha\beta}(\phi)\xi^\alpha\xi^\beta+\ldots\,, \quad W_{\alpha\beta}=W_{\beta\alpha}\,, \quad
W_{\alpha\beta}\approx\Gamma_\alpha^i\Gamma_\beta^j\partial^2_{ij}S\,.
\end{equation}
The similar procedure applies to iteratively solving order by order the
identities for identities proceeding from zero order (\ref{omega}),
(\ref{eta}).

Given the regularity of the gauge algebra of the involutive closure
(\ref{Closure}) described at the beginning of this section, no
obstructions can arise to the iterative inclusion of the
Stueckelberg fields at any order. This can be proven by the tools of
homological perturbation theory as described for the irreducible
case in the article \cite{Lyakhovich:2021lzy}. The main distinction
of this proof from the usual homological perturbation theory
procedures of gauge theories \cite{Henneaux:1992ig} is the unusual
grading, where positive resolution degree is assigned to the
Stueckelberg fields and their anti-fields, unlike the other fields.
The aspect of reducibility can be accounted for in the homological
perturbation theory with this grading in a natural way. This issue
will be addressed elsewhere. From the point of view of the
application in specific models, only the fact is important that the
described procedure for including the Stueckelberg fields is
unobstructed at all iteration steps.

\section{Reducible Stueckelberg symmetry and dual formulation for massive spin 1.}
In this Section, we exemplify the general method of inclusion
of Stueckelberg fields with reducible gauge symmetry by the case of
Proca model. The usual Stueckelberg scalar corresponds to the
completion of the Proca system by the first order
consequence --- transversaliity condition. This is sufficient to
make the Proca system involutive. However, the system can be
completed also by the third order consequences, and it remains
involutive. This option of the third order involutive closure, being
treated by the procedure of previous section, leads to inclusion of
the antisymmetric second rank tensor as the Stueckelberg field. The
third order consequences turn out obeying the gauge identities of
their own (cf. (\ref{Ztau})), so we arrive at reducible
Stueckelberg symmetry. This is no surprise once the antisymmetric
tensor is introduced. The Stueckelberg action includes four
derivatives, while the theory remains equivalent to the original
Proca system. Besides exemplifying the general method, this case may
have some interest of its own, as it demonstrates the scheme for
constructing dual formulations for the fields of the same
spin.

Consider the Proca Lagrangian for massive vector field $A_\mu$ in $d=4$
Minkowski space,
\begin{equation}\label{ProcaL}
\displaystyle
\mathcal{L}=-\frac{1}{4}F_{\mu\nu}F^{\mu\nu}+\frac{m^2}{2}A_\mu
A^\mu, \quad F_{\mu\nu}=\partial_\mu A_\nu-\partial_\nu A_\mu \, .
\end{equation}
The Proca equations
\begin{equation}\label{ProcaEq}
\displaystyle \frac{\delta S}{\delta A^\mu}\equiv\square
A_\mu-\partial_\mu\partial^\nu A_\nu+m^2A_\mu=0
\end{equation}
are not involutive as such, as they admit the first order
differential consequence
\begin{equation}\label{divA}
\displaystyle \tau\equiv\partial^\mu\frac{\delta S}{\delta
A^\mu}=m^2\partial^\mu A_\mu \, .
\end{equation}
The system (\ref{ProcaEq}), (\ref{divA}) is involutive, so it can
serve as the starting point for inclusion Stueckelberg fields. Once
the consequence (\ref{divA}) is a scalar, corresponding Stueckelberg
field should be scalar. The procedure of previous section for
inclusion of Stueckeleberg fields, being applied to the system
(\ref{ProcaEq}), (\ref{divA}), reproduces the usual Stueckelberg
formulation for the massive spin 1. The system (\ref{ProcaEq}),
(\ref{divA}) can be complemented by the third order consequences,
and still remains involutive. Consider the differential consequences
\begin{equation}\label{3tau}
\displaystyle
\tau_{\mu\nu}\equiv\frac{1}{2}(\partial_\mu\delta_\nu^\rho-\partial_\nu\delta_\mu^\rho)\frac{\delta
S}{\delta A^\rho}=\frac{1}{2}(\square+m^2)F_{\mu\nu} \, .
\end{equation}
These equations mean that the strength tensor of original field
$A_\mu$ obeys Klein-Gordon  equation. The system (\ref{ProcaEq}),
(\ref{divA}), (\ref{3tau}), being equivalent to the original Proca
equations, is also involutive. So, it can be another starting point
for including Stueckelberg fields. Following the general scheme of
the previous section, let us list the identities between the
equations of the involutive system (\ref{ProcaEq}), (\ref{divA}),
(\ref{3tau}). First, there are the identities (\ref{GI})  that
follow from the definitions of the consequences. For the involutive
closure of Proca system (\ref{ProcaEq}), (\ref{divA}), (\ref{3tau}),
these identities read
\begin{equation}\label{ProcaGI1}
 -\partial^\mu\frac{\delta S}{\delta A^\mu}+\tau=0\, ;
\end{equation}
\begin{equation}\label{ProcaGI2}
-\frac{1}{2}(\partial_\mu\delta_\nu^\rho-\partial_\nu\delta_\mu^\rho)\frac{\delta
S}{\delta A^\rho}+\tau_{\mu\nu}=0 \, .
\end{equation}
The consequences $\tau_{\mu\nu}$ (\ref{3tau}) are reducible in the
sense of identity (\ref{Ztau}). This identity reads
\begin{equation}\label{ProcaZT}
\displaystyle
\varepsilon^{\mu\nu\rho\lambda}\partial_\nu\tau_{\rho\lambda}=0\, ,
\end{equation}
where $\varepsilon^{\mu\nu\rho\lambda}$ is Levi-Chivita symbol.
These identities are reducible in their own turn, as the divergence
of the l.h.s. identically vanishes for any $\tau_{\mu\nu}$. It is
the second level identity (cf. (\ref{Z1})):
\begin{equation}\label{ProcaZ1}
\displaystyle
\partial_\mu\varepsilon^{\mu\nu\rho\lambda}\partial_\nu=0 \, .
\end{equation}
The identities (\ref{ProcaGI1}), (\ref{ProcaGI2}) between the
equations of the third order involutive closure of Proca system and
the identities of identities (\ref{ProcaZT}), (\ref{ProcaZ1}) allow
one to identify all the ingredients needed for inclusion of
Stueckelberg fields with reducible gauge symmetry: the generators of
consequences $\Gamma$ (\ref{tau}), null-vectors $Z$ of $\Gamma$'s
(cf.(\ref{Z})), and null-vectors of the null-vectors $Z_1$ (cf.
(\ref{Z1})):
\begin{equation}\label{ProcaGZZ}
\Gamma^\mu=-\partial^\mu, \quad \Gamma_{\mu\nu}{}^\rho=-\frac{1}{2}(\partial_\mu\delta_\nu^\rho-\partial_\nu\delta_\mu^\rho), \quad Z^{\mu\lambda\rho}=\varepsilon^{\mu\nu\lambda\rho}\partial_\nu, \quad Z_{1\mu}=\partial_\mu\,.
\end{equation}
With all the ingredients at hands, following the general procedure
of Section 2, we iteratively construct the Stueckelberg action,
generators of gauge symmetries, and symmetries of symmetries. Once
the original action is quadratic and the identity generators
(\ref{ProcaGZZ}) are field-independent, the procedure terminates at
the first iteration. The Stueckelberg action and reducible gauge
symmetry transformations read
\begin{equation}\label{ProcaS-St}
\begin{array}{c}
\displaystyle \mathcal{S}_{St}=\int d^4x \Big(-\frac{1}{2}\partial_\mu A_\nu F^{\mu\nu}-\frac{1}{2}\partial_\mu\partial^\rho B_{\nu\rho}\big(\partial^\mu\partial_\lambda B^{\nu\lambda}+2\,\partial^\mu A^\nu\big)\\[3mm]
\displaystyle +\,\frac{m^2}{2}\big(A_\mu
A^\mu+\partial_\mu\varphi\,\partial^\mu\varphi+\partial^\nu
B_{\mu\nu}\partial_\rho
B^{\mu\rho}\big)+m^2A_\mu\big(\partial^\mu\varphi+\partial_\nu
B^{\mu\nu}\big)\Big) \, ,
\end{array}
\end{equation}
\begin{equation}\label{Proca-GT}
\delta_\epsilon
A^\mu=-\partial^\mu\epsilon-\partial_\nu\epsilon^{\mu\nu},\qquad
\delta_\epsilon \varphi=\epsilon, \qquad\delta_\epsilon
B^{\mu\nu}=\epsilon^{\mu\nu}+\varepsilon^{\mu\nu\rho\lambda}\partial_\rho\epsilon_\lambda
\, ,
\end{equation}
where $\varphi$ and $B^{\mu\nu}=-B^{\nu\mu}$ are the Stueckelberg
fields corresponding to the consequences $\tau$ and $\tau_{\mu\nu}$
(\ref{divA}), (\ref{3tau}), while $\epsilon$, $\epsilon_\mu$,
$\epsilon_{\mu\nu}$ are the gauge parameters corresponding to the
gauge identities (\ref{ProcaGI1}), (\ref{ProcaGI2}),
(\ref{ProcaZT}). By direct computation, one can easily see that
action (\ref{ProcaS-St}) enjoys symmetry (\ref{Proca-GT}) indeed.
Given the null-vectors $Z$ and $Z_1$ (\ref{ProcaGZZ}), the
symmetries of symmetries are constructed following the general
prescription (\ref{omega}), (\ref{eta}),
\begin{equation}\label{ProcaO}
 \delta_\omega\epsilon=0\,, \qquad
 \delta_\omega\epsilon_\mu=-\omega_\mu-\partial_\mu\omega\, , \qquad
\delta_\omega\epsilon^{\mu\nu}=\varepsilon^{\mu\nu\rho\lambda}\partial_\rho\omega_\lambda\,,
\end{equation}
\begin{equation}\label{ProcaEta}
\delta_\eta\omega=\eta \,
,\qquad\delta_\eta\omega_\mu=-\partial_\mu\eta\, ,
\end{equation}
where (\ref{ProcaO}) are the gauge symmetry transformations of the
original gauge parameters $\epsilon$, while (\ref{ProcaEta}) is the
gauge symmetry of the second level gauge parameters $\omega$.

Consider the Lagrangian equations for Stueckelberg action (\ref{ProcaL}),
\begin{equation}\label{Proca-St-E1}
\displaystyle \frac{\delta \mathcal{S}_{St}}{\delta
A^\mu}\equiv\square A_\mu-\partial_\mu\partial^\nu
A_\nu+\square\partial^\nu
B_{\mu\nu}+m^2A_\mu+m^2\partial_\mu\varphi+m^2\partial^\nu
B_{\mu\nu}=0\, ,
\end{equation}
\begin{equation}\label{Proca-St-E2}
\displaystyle \frac{\delta \mathcal{S}_{St}}{\delta
\varphi}\equiv-m^2(\square\varphi+\partial^\mu A_\mu)=0 \, ,
\end{equation}
\begin{equation}\label{Proca-St-E3}
\displaystyle \frac{\delta \mathcal{S}_{St}}{\delta
B^{\mu\nu}}\equiv\frac{1}{2}(\square+m^2)\big(\partial_\mu\partial^\rho
B_{\rho\nu}-\partial_\nu\partial^\rho B_{\rho\mu}+
\partial_\mu A_\nu-\partial_\nu A_\mu\big)=0 \, .
\end{equation}
These equations involve the fourth order derivatives, so equivalence
with the original Proca theory may seem doubtful. However, these
equations enjoy the reducible gauge symmetry (\ref{Proca-GT}). This
symmetry admits gauge fixing conditions
\begin{equation}\label{ProcaGF0}
\varphi=0\,, \quad B_{\mu\nu}=0 \, .
\end{equation}
This gauge eliminates all the Stueckelberg fields and reduces the
system to Proca equations (\ref{ProcaEq}).

It is interesting to notice another admissible gauge fixing for the
symmetry (\ref{Proca-GT}):
\begin{equation}\label{ProcaGF1}
\varphi=0\,, \qquad A_\mu=0\,, \qquad
\varepsilon_{\mu\nu\rho\lambda}\partial^\nu B^{\rho\lambda}=0\, .
\end{equation}
As this gauge fixing kills scalar $\varphi$ and vector field $A_\mu$, equations (\ref{Proca-St-E1})-(\ref{Proca-St-E2}) reduce to third-order equation (\ref{B}), while (\ref{Proca-St-E3}) becomes its differential consequence.
Let us detail fixing of the gauge parameters by conditions
(\ref{ProcaGF1}). Taking variation of (\ref{ProcaGF1}) we arrive at
the conditions
\begin{equation}\label{GFvar}
\delta_\epsilon \varphi=\epsilon=0\, ,\quad \delta_\epsilon
A_\mu=\partial_\mu\epsilon +\partial^\nu\epsilon_{\mu\nu}=0\, ,
\quad \varepsilon_{\mu\nu\rho\lambda}\partial^\nu \delta_\epsilon
B^{\rho\lambda}=
\varepsilon_{\mu\nu\rho\lambda}\partial^\nu\epsilon^{\rho\lambda}-\square\epsilon_\mu+\partial_\mu\partial^\nu\epsilon_\nu=0
\,.
\end{equation}
So, the gauge conditions (\ref{ProcaGF1}) restrict the gauge
parameters by the relations
\begin{equation}\label{GPE}
\epsilon=0\,,\quad \partial_\mu\epsilon^{\mu\nu}=0, \quad
\varepsilon_{\mu\nu\rho\lambda}\partial^\nu\epsilon^{\rho\lambda}-\square\epsilon_\mu+\partial_\mu\partial^\nu\epsilon_\nu=0
\, .
\end{equation}
Once $\epsilon=0$, the second of these equations means
$\epsilon^{\mu\nu}=\varepsilon^{\mu\nu\lambda\rho}\partial_\lambda\omega_\rho$,
where $\omega_\lambda$ is arbitrary. Substituting that into the last
relation we see that the difference between the gauge parameter
$\epsilon_\mu$ and $\omega_\mu$ obeys free Maxwell equations.
Maxwell equations have unique solution modulo the gradient of
arbitrary scalar  $\partial_\mu \omega$, given the Cauchy data. So,
the general solution of equations (\ref{GPE}) reads
\begin{equation}\label{epsilon-omega}
    \epsilon= 0 \, \qquad \epsilon_\mu =  \omega_\mu + \partial_\mu\omega, \quad
    \epsilon^{\mu\nu} = \varepsilon^{\mu\nu\rho\lambda}\partial_\rho\omega_\lambda\, ,
\end{equation}
where $\omega_\mu, \omega$ are arbitrary functions. This means, the
gauge conditions (\ref{ProcaGF1}) fix parameters
$\epsilon,\epsilon_\mu, \epsilon_{\mu\nu}$ modulo symmetry of
symmetry (\ref{ProcaO}). The ambiguity of this type always remains
unfixed at the level of field equations for original fields in the
case of reducible gauge symmetry. In the BRST formalism, this
ambiguity is fixed by imposing gauge conditions on the ghosts and
introducing ghosts for ghosts \cite{Henneaux:1992ig}.

Admissibility of the gauge fixing condition such that kills the
original vector field means that $A_\mu$ can be considered as a pure
gauge from the viewpoint of the action (\ref{ProcaL}) with gauge
symmetry (\ref{Proca-GT}). This is true indeed, given the
transformation $\delta_\epsilon A_\mu$ (\ref{Proca-GT}) which
demonstrates that both gradient and transverse parts are ambiguous
of the vector $A_\mu$, so only zero modes can survive in the gauge
transformations. Once transformations for the fields $A_\mu$ and
$B^{\mu\nu}$ share the same gauge parameter $\epsilon^{\mu\nu}$, the
gauge ambiguity can be equally well fixed either by the conditions
killing $A$ and residual ambiguity in $B$, or by fixing $B$.

The equations (\ref{B}), being one of the gauge fixed forms of the
Stueckelberg system (\ref{Proca-St-E1}), (\ref{Proca-St-E2}),
(\ref{Proca-St-E3}) are equivalent to the original Proca system.
These third order non-Lagrangian equations can be considered a dual
form of the vector representation (\ref{KG}) of massive spin 1
particle,  as it has been already explained in the introduction. One
can switch between these dual forms by imposing different gauges in
the same Lagrangian theory. This example demonstrates that if the
inclusion of Stueckelberg fields begins with the higher order
involutive closure of the original theory, the Stueckelberg action,
being equivalent to the original non-involutive theory, can include
dual formulations of the same irreps. This topic is further
discussed in Conclusion.

\section{Conclusion}
Let us summarize and discuss the results. First, we propose a
systematic way for inclusion of Stueckelberg fields such that
guarantees equivalence of the resulting gauge theory to the original
system. The starting point for inclusion of Stueckelberg fields is
the involutive closure of original Lagrangian equations
(\ref{Closure}). If the closure includes an over-complete set of
consequences (see (\ref{Ztau})), the Stueckelberg symmetry turns out
reducible. In any case, the Stueckelberg theory is iteratively
constructed for any involutive closure of Lagrangian equations
without obstructions at any stage, be the consequences (\ref{tau})
reducible or not. In this sense, the covariant method is a complete
analogue to the Hamiltonian method of conversion of the second class
constraints into the first class ones.

The interesting option for inclusion of Stueckelberg fields is to
start with the involutive closure of the higher order than it is
minimally sufficient. This option is exemplified in Section 3 by the
third order involutive closure of Proca model, where the added
consequences are reducible. Following the general procedure of
inclusion of the Stueckelberg fields, we arrive to the higher
derivative Stueckelberg action (\ref{ProcaS-St}) which is equivalent
to the first derivative Proca action. This Stueckelberg model for
massive spin 1 turns out comprising two dual field theoretical
realizations for the same irreducible representation. The first one
is the original Proca model, and the second one is the third order
formulation (\ref{B}) in terms of the antisymmetric tensor field.
The field $B^{\mu\nu}$ can be considered as a potential for the
original transverse vector (cf. (\ref{Transverse})). Notice that
various dual formulations are studied once and again for the same
spin representation. For the most recent results on this topic and
further references we refer to the article \cite{Boulanger:2020yib}.
Important motivation for studying dual formulations is that they are
inequivalent, in general, w.r.t. inclusion of interactions. Among
the examples of this sort, we can mention the representation of the
massless spin 2 by the third rank tensor field with Young diagram of
the hook type \cite{Curtright:1980yk}. Unlike the representation of
the same spin by the symmetric second rank tensor, the hook does not
admit inclusion of consistent interactions \cite{Bekaert:2002uh}.
Similar phenomena are observed among the higher spin gravities. In
particular, the long known light-cone analysis of the higher spin
vertices in Minkowski space \cite{Metsaev:1991mt} demonstrates
admissibility of the interactions such that are missing among the
deformations of Fronsdal's Lagrangians for symmetric tensors. There
is a growing evidence that Lagrangians for dual formulations of
higher spins can admit these vertices. For the recent results,
discussion of the area, and further references we refer to
\cite{Conde:2016izb}, \cite{Krasnov:2021nsq}. Notice that the
considered dual formulations are typically connected to each other
algebraically, hence all the actions are of the same order. Proposed
scheme of inclusion Stueckelberg fields proceeds from the involutive
closure of the original Lagrangian equations. If the starting point
is the higher order closure of the original system, corresponding
Stueckelberg field, being candidate for the dual to the original
field, would be involved in the Lagrangian with higher derivatives.
This dual would be connected to the original field by a differential
relation, like a potential (cf. (\ref{Transverse})). So, this
scenario of inclusion Stueckelberg fields can serve as a tool for
constructing a different type of dual formulations. For example, if
the original fields are symmetric, the second order Lagrangian
equations can admit the third order involutive closure with
differential consequences, being the tensors of hook type.
Corresponding higher derivative Stueckelberg Lagrangian has to be
equivalent to the original one by construction, while the hook
tensors would serve as dual to the original fields, following the
pattern of Section 3. These dual models can have their chances for
consistent interactions as the potentials can be less obstructive to
deformations than corresponding strength tensors.

\subsection*{Acknowledgements} We thank  A.~Sharapov for valuable discussions.
The work is supported by the Foundation for the Advancement of
Theoretical Physics and Mathematics ``BASIS".


\begin{thebibliography}{11}
 \bibitem{Stueckelberg} St\"uckelberg, Ernst C.G. (1938). "Die Wechselwirkungskr\"afte in der Elektrodynamik und in der Feldtheorie der Kr\"afte". Helvetica Physica Acta (in German). \textbf{11}: 225
\bibitem{Ruegg:2003ps}
H.~Ruegg and M.~Ruiz-Altaba, ``The Stueckelberg field'', Int. J.
Mod. Phys. A \textbf{19} (2004), 3265-3348
 [arXiv:hep-th/0304245].
\bibitem{Boulanger:2018dau}
N.~Boulanger, C.~Deffayet, S.~Garcia-Saenz and L.~Traina,
``Consistent deformations of free massive field theories in the
Stueckelberg formulation'', JHEP \textbf{07}, 021 (2018)
[arXiv:1806.04695 [hep-th]].
\bibitem{Faddeev:1986pc}
L.~D.~Faddeev and S.~L.~Shatashvili, ``Realization of the Schwinger
Term in the Gauss Law and the Possibility of Correct Quantization of
a Theory with Anomalies'', Phys. Lett. B \textbf{167} (1986),
225-228

\bibitem{Batalin:1986aq}
I.~A.~Batalin and E.~S.~Fradkin, ``Operator Quantization of
Dynamical Systems With Irreducible First and Second Class
Constraints'', Phys. Lett. B \textbf{180} (1986), 157-162 [erratum:
Phys. Lett. B \textbf{236} (1990), 528]

\bibitem{Batalin:1991jm}
  I.~A.~Batalin and I.~V.~Tyutin,
``Existence theorem for the effective gauge algebra in the
generalized canonical formalism with Abelian conversion of second
class constraints'',
  Int.\ J.\ Mod.\ Phys.\ A {\bf 6} (1991) 3255.

\bibitem{Batalin:2005df}
  I.~Batalin, M.~Grigoriev and S.~Lyakhovich,
  ``Non-Abelian conversion and quantization of non-scalar second-class constraints'',
  J.\ Math.\ Phys.\  \textbf{ 46} (2005) 072301
  [arXiv:hep-th/0501097].

\bibitem{Lyakhovich:2021lzy}
S.~L.~Lyakhovich, ``General method for including Stueckelberg
fields'', Eur. Phys. J. C {\bf  81}, 472 (2021) [arXiv:2102.10579
[hep-th]].

\bibitem{Involution}
  D.~S.~Kaparulin, S.~L.~Lyakhovich and A.~A.~Sharapov,
  ``Consistent interactions and involution'',
  JHEP  \textbf{1301} (2013) 097
  [arXiv:1210.6821 [hep-th]].

\bibitem{Lyakhovich:2004xd}
S.~L.~Lyakhovich and A.~A.~Sharapov, ``BRST theory without
Hamiltonian and Lagrangian'', JHEP \textbf{03} (2005), 011
[arXiv:hep-th/0411247].

\bibitem{Kazinski:2005eb}
P.~O.~Kazinski, S.~L.~Lyakhovich and A.~A.~Sharapov, ``Lagrange
structure and quantization'', JHEP \textbf{07} (2005), 076
[arXiv:hep-th/0506093].

\bibitem{Henneaux:1992ig}
  M.~Henneaux and C.~Teitelboim,
  ``Quantization of gauge systems'',
  Princeton, USA: Univ. Pr. (1992) 520 p.

\bibitem{Boulanger:2020yib}
N.~Boulanger and V.~Lekeu, ``Higher spins from exotic
dualisations'', JHEP \textbf{03} (2021), 171 [arXiv:2012.11356
[hep-th]].

\bibitem{Curtright:1980yk}
T.~Curtright, ``Generalized gauge fields'', Phys. Lett. B
\textbf{165} (1985), 304.

\bibitem{Bekaert:2002uh}
X.~Bekaert, N.~Boulanger and M.~Henneaux, ``Consistent deformations of dual formulations of linearized gravity: A No go result'', Phys. Rev. D \textbf{67} (2003), 044010 [arXiv:hep-th/0210278].

\bibitem{Metsaev:1991mt}
R.~R.~Metsaev, ``Poincare invariant dynamics of massless higher
spins: Fourth order analysis on mass shell'', Mod. Phys. Lett.
\textbf{A6} (1991), 359--367;\, R.~R.~Metsaev, ``S matrix approach
to massless higher spins theory. 2: The Case of internal symmetry'',
Mod. Phys. Lett. \textbf{A6} (1991), 2411--2421.

\bibitem{Conde:2016izb} E.~Conde, E.~Joung, and K.~Mkrtchyan,
``Spinor-Helicity Three-Point Amplitudes from Local Cubic
Interactions'', JHEP \textbf{08} (2016), 040 [arXiv:1605.07402
[hep-th]].

\bibitem{Krasnov:2021nsq}
K.~Krasnov, E.~Skvortsov and T.~Tran, ``Actions for Self-dual Higher
Spin Gravities'', [arXiv:2105.12782 [hep-th]].

\end{thebibliography}
\end{document}